\documentclass[APS,twocolumn,superscriptaddress]{revtex4-1}%

\usepackage{amsmath,amssymb}
\usepackage{capt-of}
\usepackage{color}
\usepackage{float}
\usepackage{gensymb} 
\usepackage{glossaries}
\usepackage{graphicx}
\usepackage[
    unicode,
    pdfauthor={Canales et al.},
    colorlinks=true,
    urlcolor=blue,
    linkcolor=blue,
    citecolor=blue
]{hyperref}
\usepackage{lmodern}
\usepackage[version=3]{mhchem}
\usepackage{multirow}
\usepackage{varwidth}
\usepackage{textcomp}
\usepackage{todonotes}
\usepackage{units}
\usepackage{soul}

\newcommand{\eps}{\varepsilon}

\usepackage{xr}

\makeatletter
\newcommand*{\addFileDependency}[1]{
  \typeout{(#1)}
  \@addtofilelist{#1}
  \IfFileExists{#1}{}{\typeout{No file #1.}}
}
\makeatother

\newcommand*{\myexternaldocument}[1]{%
    \externaldocument{#1}%
    \addFileDependency{#1.tex}%
    \addFileDependency{#1.aux}%
}
\myexternaldocument{SM_water}

\begin{document}
\title{Self-hybridized vibrational-Mie polaritons in water droplets}

\newcommand{\phys}{
    Department of Physics,
    Chalmers University of Technology,
    412~96, G\"oteborg, Sweden
}

\author{Adriana Canales}
\affiliation{\phys}

\author{Oleg V. Kotov}
\affiliation{\phys}

\author{Betül Küçüköz}
\affiliation{\phys}

\author{Timur O. Shegai}
\email{timurs@chalmers.se}
\affiliation{\phys}


\begin{abstract}  

We study the self-hybridization between Mie modes supported by water droplets with stretching and bending vibrations in water molecules. Droplets with radii $>2.7~\mu m$ are found to be polaritonic on the onset of the ultrastrong light-matter coupling regime. Similarly, the effect is observed in larger deuterated water droplets at lower frequencies. Our results indicate that polaritonic states are ubiquitous in nature and occur in water droplets in mists, fogs, and clouds. This finding may have implications not only for polaritonic physics but also for aerosol and atmospheric sciences.


\end{abstract}
\date{\today}

\maketitle

Water, one of the most widespread compounds on Earth, plays a vital role in a broad range of physical, chemical, biological, geological, atmospheric, and climate-related phenomena~\cite{stillinger1980water}. While being one of the most studied substances, water encompasses a number of unusual properties (e.g. triple point, anomalous density, anomalous heat expansion, etc.), some of which keep puzzling researchers~\cite{kim2020experimental,yang2021direct}. Concurrently, advances in the realm of strong light-matter coupling have significantly impacted several research fields, including exciton transport~\cite{feist2015extraordinary,schachenmayer2015cavity,balasubrahmaniyam2023enhanced}, resonance energy transfer~\cite{coles2014polariton}, photochemistry~\cite{hutchison2012modifying,Munkhbat2018,peters2019control}, charge transport~\cite{orgiu2015conductivity}, and ground-state chemical reactivity~\cite{thomas2019tilting,ahn2023modification}. Central to these developments are polaritons -- hybrid states of light and matter, which arise as a result of strong coupling between photonic modes and material's electronic or vibrational excitations~\cite{Ebbesen2021Manipulating}. Water is a promising polaritonic material platform, owing to its high oscillator strength manifested in pronounced absorption, particularly in the mid-infrared range~\cite{Fiedler2020permittivity}. Indeed, vibrational strong coupling between water molecules and planar microcavities has been demonstrated in several recent works~\cite{vergauwe2019modification,li2020cavity,lather2020improving,HiraiVSCCrystal,imperatore2021reproducibility,fukushima2021vibrational,fukushima2022inherent}. Self-assembled microcavities and polaritons have also been recently observed in aqueous solutions~\cite{munkhbat2021tunable}. These realizations, however, required an external cavity, typically comprised of two metallic mirrors.



To overcome the limitations of optical cavities based on metallic mirrors, \textit{self-hybridized} polaritons have arisen recently. In these polaritons, optical modes supported by (nano)structured material's geometry hybridize with electronic or vibrational excitations of the very same material~\cite{SC-canales2021abundance}. These polaritons, inspired by the original work of Hopfield on bulk polaritons~\cite{hopfield1958theory}, have been experimentally and theoretically studied in various systems, including excitonic slabs~\cite{SC-munkhbat2018self,SC-gogna2020selfReS2,SC-thomas2021cavityFree,Canales2023perfect,SC-dirnberger2023magneto}, nanodisks~\cite{SC-verre2019transition}, nanotubes~\cite{SC-pandya2021self}, and microspheres~\cite{Platts2009wgm}. However, these observations were made in solid-state platforms. The potential existence of polaritons in liquids, and in water droplets in particular, has so far been shown only theoretically~\cite{SC-canales2021abundance}.

There are two fundamental advantages associated with studying polaritons within liquid droplets. First, the spherical geometry permits the exact solution of the electromagnetic eigenstate problem by means of Mie theory. Second, the production of droplets is a spontaneous process driven by the high surface tension of the liquid, resulting in nearly ideal spherical shapes. The latter renders water droplets as promising whispering-gallery mode resonators, characterized by quality factors exceeding 10$^7$ \cite{maayani2016water,giorgini2017fundamental}, and as highly efficient Mie scatterers when subjected to optical levitation~\cite{Ward2008Mie,Marmolejo2023Fano}. Moreover, a detailed knowledge of electromagnetic eigenstates of water droplets is essential for an improved understanding of liquid water's properties. These eigenstates determine the spectroscopic characteristics of water droplets in the mid-infrared range, which may impact scientific disciplines beyond polaritonics. These findings could, for instance, find implications in aerosol and atmospheric research, where previous measurements of water droplets in their clustered forms (e.g. mists, fogs, clouds) have shown spectroscopic peculiarities in the mid-infrared range around molecular vibrations~\cite{Blau1966CloudSpec,Arnott1997cloud}. However, these peculiarities have not been previously interpreted in terms of vibrational-Mie self-hybridization. 



Here, we aim to study polaritonic eigenmodes of water droplets by means of Mie theory and infrared spectroscopy. We focus on measuring the optical density of water droplet ensembles, comprising particles of various sizes, typically within a range of a few microns. To produce such a laboratory mist, we use a mesh nebulizer. The infrared spectroscopy data, in combination with Mie theory, reveals the polaritonic nature of the observed signal and the size distribution of droplets present in the samples. Furthermore, we conducted a similar experiment with deuterated (heavy) water and observed spectral shifts due to modified molecular vibrations and altered droplet sizes. These findings ensure that water droplets exhibit polaritonic self-hybridization between Mie modes of the spheres and molecular vibrations on the onset of ultrastrong light-matter coupling.


\begin{figure}[ht!]
\includegraphics[width=0.45\textwidth]{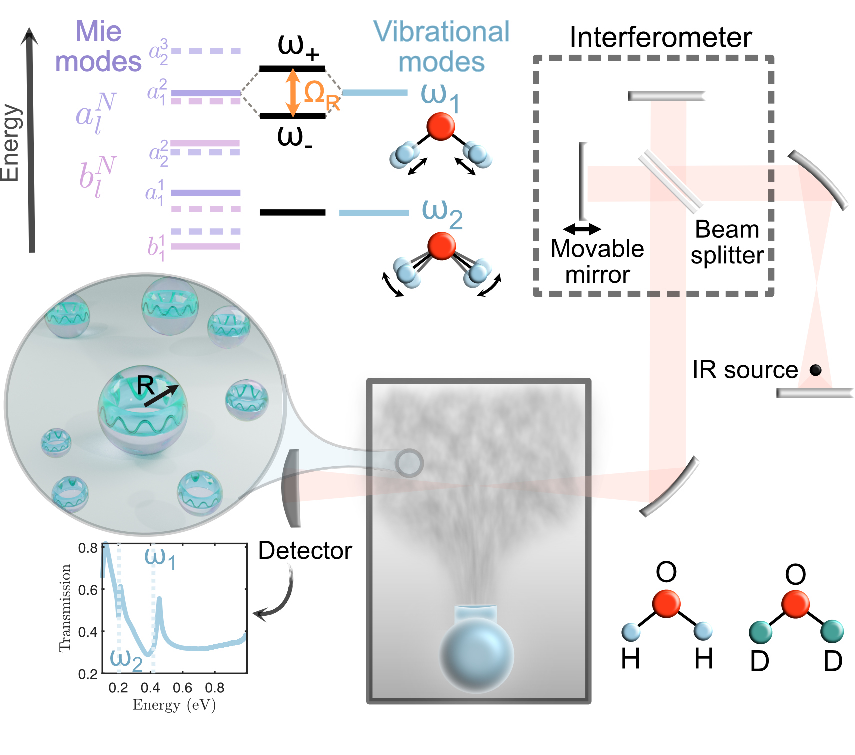}
\caption{\footnotesize{\textbf{Concept of experiment on self-hybridization between Mie modes and molecular vibrations in water droplets}. A vibrating mesh nebulizer generates a laboratory mist with droplet radii $<$ 5~$\mu$m. Infrared spectroscopy in transmission mode reveals Mie modes strongly coupled to molecular vibrations within water droplets, which is manifested in the emergence of polaritonic energy levels, $\hbar\omega_{\pm}$, separated by Rabi splitting, $\Omega_R$. The experiments were performed with both regular (H$_2$O) and heavy (D$_2$O) water at different energy and droplet size ranges.}}
\label{Fig1.Concept}
\end{figure}


The laboratory mist used in the experiment was created using a vibrating mesh nebulizer (Evolu Air Pro) calibrated to produce droplets smaller than 5 $\mu$m radius with 0.9\% saline water (Figure S1). As depicted in \autoref{Fig1.Concept}, the laboratory mist was set on the optical path of a Fourier transform infrared spectrometer (FTIR - Bruker Vertex 70v) in transmission mode, covering the spectral range of 0.1 to 1 eV (1.2 - 12.4 $\mu$m). The optical density ($OD$) was then calculated as $OD = - \log{(T)}$. As we will see later, the $OD$ can be calculated via the extinction cross-section of individual droplets using Beer-Lambert's law. 

Water droplets forming the mist sustain Mie modes that interact with the symmetric and anti-symmetric O-H stretching as well as the H-O-H bending vibrations. To demonstrate their polaritonic nature, we look for the main signature of strong coupling: the mode Rabi splitting ($\Omega_R$) at zero detuning, which is the energy difference between upper ($\hbar\omega_+$) and lower ($\hbar\omega_-$) polariton branches, as shown in~\autoref{Fig1.Concept} (orange arrow).

\begin{figure}[t!]
\includegraphics[width=0.5\textwidth]{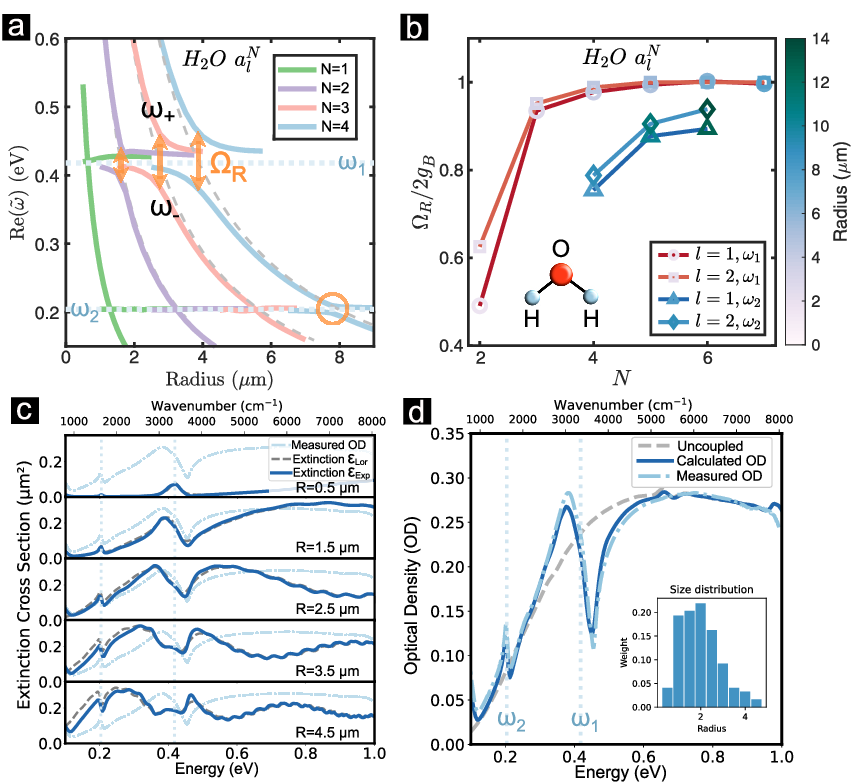}
\caption{\footnotesize{\textbf{Polaritons in H$_{2}$O droplets}. (a) Eigenfrequencies of $a_{l=1}^N$ as a function of droplet radius. In a certain size range, the mode splittings are observed. The Rabi splitting is marked with orange arrows. (b) The Rabi splitting increases with $N$ and converges to the bulk Rabi splitting $2g_B$. (c) Calculated extinction cross-sections of individual water droplets as a function of droplet size. The cross-sections are calculated using experimental (Lorentz) permittivity shown in blue (dashed) lines. The uncoupled molecular vibrations are marked with vertical dotted lines. (d) Experimental and fitted optical density using the droplet size distribution shown in the inset. The grey dashed line shows calculated $OD$ for the same size distribution but with vibrational modes artificially switched off, thus representing \emph{uncoupled} spheres. Vertical dotted lines mark the position of regular water vibrations.
}}
\label{Fig2.H2O}
\end{figure}

Measuring the Rabi splitting in planar microcavities is straightforward because, in a typical situation, a spectrally isolated optical mode(s) interacts with a single molecular vibration. In a mist containing droplets of various sizes, however, the $OD$ aggregates the contributions of all Mie modes within individual droplets and furthermore accounts for the entire size distribution (\autoref{Fig1.Concept}), thus hindering direct observation of the Rabi splitting. Therefore, it is important first to calculate the modes resulting from the interaction between vibrational modes and single Mie modes in a sphere (\autoref{Fig2.H2O}a). Then, the extinction spectrum of a single droplet with a fixed radius can be evaluated by summing all modes, as exemplified in \autoref{Fig2.H2O}c. Finally, to account for specific size distribution, the aggregated extinction spectrum of the mist is obtained as a weighted sum of contributions from individual water droplets of varying diameters, as shown in \autoref{Fig2.H2O}d. 

We start by calculating the eigenstates of individual water droplets. A water droplet is an open system, therefore we use complex eigenfrequencies to describe its quasinormal modes, $\tilde{\omega}=\omega -i \gamma/2$~\cite{QNMs-Lalanne2018}. The real part of the complex eigenfrequency describes the resonant frequency (or energy, $\hbar \omega$), while the imaginary part describes its total decay rate, $\gamma$. 

The eigenfrequencies are calculated as poles of the scattering coefficients given by the classical Mie solution \cite{Bohren2004}. Such scattering coefficients for a water sphere in air are:
\begin{equation}\label{eq:scatt}
\begin{aligned}
        a_l = \frac{m \psi_l(m x)  \psi_l '(x) - \psi_l (x)\psi_l '(m x)}
                { m \psi_l(m x)  \xi_l '(x) - \xi_l (x)\psi_l '(m x)},\\
    b_l = \frac{\psi_l(m x)  \psi_l '(x) - m \psi_l (x)\psi_l '(m x)}
                {\psi_l(m x)  \xi_l '(x) - m \xi_l (x)\psi_l '(m x)}\;.
\end{aligned}
\end{equation}
Here, the size parameter, $x=k R$, includes light's wavevector $k$ and the radius of the droplet, $R$. We use the Ricatti-Bessel functions $\psi_l(x)=x j_l(x)$ and $\xi_l(x)=x h_l^{(1)}(x)$, where $j_l(x)$ are spherical Bessel functions of the first kind, and $h_l^{(1)}(x)$ are Hankel functions of the first kind. The relative refractive index $m = n_{water}/n_{air}$, where $n_{air}=1$ and $n_{water}=\sqrt{\eps(\omega)}$, with $\eps(\omega)$ being the water permittivity.

To analytically calculate the complex eigenfrequencies, the experimental water permittivity \cite{segelstein1981complex} was approximated by a Lorentzian function, $\varepsilon(\omega)$, using the fitting provided by Fiedler \textit{et al.} \cite{Fiedler2020permittivity} for the infrared region of interest (0.1-1 eV):
\begin{equation} \label{Eq:Lorentz}
  \eps(\omega) = \varepsilon_{\infty} + \frac{f_{1}\omega_{1}^2}{\omega_{1}^2-\omega^2 - i\gamma_{1}\omega} + \frac{f_{2}\omega_{2}^2}{\omega_{2}^2-\omega^2 - i\gamma_{2}\omega}\; 
\end{equation}
where $\varepsilon_{\infty}=1.75$ accounts for all higher energy excitations. Due to their spectral overlap, we model the symmetric and anti-symmetric O-H stretching vibrations as one effective oscillator with $\omega_{1}=0.418$ eV, a high oscillator strength $f_{1}=0.0717$, and a decay rate $\gamma_{1}=0.0341$ eV~\cite{bagchi2013watervibr}. The H-O-H bending was considered as one oscillator with $\omega_{2}=0.204$ eV, an oscillator strength of $f_{2}=0.0134$ and decay rate of $\gamma_{2}=0.0084$ eV~\cite{Fiedler2020permittivity}. 

The scattering coefficients in Eq. \eqref{eq:scatt} are associated with two types of modes: Transverse Magnetic (TM) and Transverse Electric (TE). TM (TE) eigenfrequencies are found when the denominator of the scattering coefficients $a_l$ ($b_l$) equals zero. The multipole mode number $l$ represents the number of electromagnetic field maxima around the circumference of the droplet. For a fixed angular number $l$ and radius $R$, multiple mode orders occur at different energies with varying radial numbers, $N$. This mode order specifies the maxima in the radial direction inside the droplet~\cite{Ward2008Mie}. To specify the mode in the text, we write $a_l^N$ and $b_l^N$, which is equivalent to TM$_{lN}$ and TE$_{lN}$~\cite{Platts2009wgm,Marmolejo2023Fano}.

All these modes co-exist within a single water droplet and contribute to its extinction cross-section.  For a droplet of a given radius, the extinction cross-section reads:
\begin{equation}\label{eq:ext}
         \sigma_{ext}=\frac{2 \pi}{k^2} \sum_{l=1}^{\infty} (2l+1)\Re(a_l+b_l)\;,
\end{equation}
which we evaluate numerically using MiePython module~\cite{prahl_scott_2023_8023972}.

To account for the different sizes in the mist, we use Beer-Lambert's law to represent the $OD$ as a sum of extinction cross-sections for each radius $R_i$, weighted by their respective concentrations $n_i$, $OD= L\sum_{i} n_i(R_i) \sigma_{{ext}_i}(R_i)$~\cite{Arnott1997cloud}. Here, $L$ is the optical path length through the mist. To fit the experimental $OD$, we calculated it in the form of the total normalized extinction cross-section with the fitting coefficients $w_i(R_i,L)\sim Ln_i(R_i)$ playing the role of static weights representing each droplet size:
 \begin{equation}\label{eq:OD}
         OD = \sum_{i} w_i(R_i,L) \sigma_{{ext}_i}(R_i)\;.
\end{equation}
The weights and radii in the size distribution were optimized to fit the experimental $OD$, using the log-normal distribution of mist from mesh nebulizers as a guide~\cite{McDermott2021lognormalmesh, SHARMA2022meshlognormal, sadeghVMNlognormal}. The experimental water permittivity was used in the fitting procedure~\cite{Fiedler2020permittivity}. The range of droplet radii produced by the mesh nebulizer was obtained from Evolu and is shown in Figure S1. Fitting attempts with other radii ranges were unsuccessful, as shown in Figure S7.



For simplicity, we start by analyzing $a_1^N$ for different radii in \autoref{Fig2.H2O}a. The real part of the eigenfrequencies is shown with colored lines for the first four radial numbers. The first radial mode (green) is weakly coupled to both water vibrations since the modes cross at zero detuning. The points of zero detuning occur where the uncoupled Mie modes (grey dashed lines) cross with the uncoupled H$_2$O vibrational modes (blue dotted lines). Droplets with radii below 1.5 $\mu$m only support the lowest TM and TE modes, resulting in weak coupling for all such droplets. See Figure S2 for the eigenfrequencies of other modes.

However, larger droplets can support higher $N$ modes, resulting in mode splitting with $\omega_1$ for $N\geq2$ and $\omega_2$ for $N\geq4$ (\autoref{Fig2.H2O}a). The Rabi splitting is marked with orange arrows. The onset of strong coupling is found when $\Omega_R>\gamma_{avg} = \frac{(\gamma_M + \gamma_v)}{2}$~\cite{torma2014strong}. Therefore, despite the visible splitting with $\omega_2$ for $N=4$ circled in \autoref{Fig2.H2O}a, it is only strongly coupled after $N=6$ (see Figure S4). On the other hand, the splitting with $\omega_1$ is enough to be strongly coupled for $N\geq3$, which corresponds to water droplets being strongly coupled for radii above 2.7 $\mu$m.

\autoref{Fig2.H2O}a displays an increase in Rabi splitting with higher $N$ due to the decay rate of the uncoupled Mie mode getting closer to the uncoupled vibrational mode, visible in the complex frequency plane (see Figure S3). This results in a reduced $\delta\gamma=\gamma_M - \gamma_v$, which increases the apparent Rabi splitting since $\Omega_R=2\sqrt{g^2 - (\delta\gamma/4)^2}$. The same pattern occurs for higher $l$, as summarized in \autoref{Fig2.H2O}b. 

Despite the observed increase of the Rabi splitting for higher-order modes, it eventually saturates as depicted in~\autoref{Fig2.H2O}b, with the ultimate limit set by the bulk Rabi splitting~\cite{SC-canales2021abundance,Platts2009wgm}, $2g_B=\omega_{1}\sqrt{f/\eps_{\infty}}\approx 85$ meV. In agreement with previous studies~\cite{fukushima2021vibrational,SC-canales2021abundance}, the system reaches the onset of the ultrastrong coupling (USC) regime as shown in Figure S6.

For a fixed radius, all hybridized modes contribute to light scattering and absorption, collectively yielding a water droplet's extinction cross-section described by Eq. \eqref{eq:ext}. The smallest droplets investigated in \autoref{Fig2.H2O}c are weakly coupled, as evidenced by mode crossing. On the other hand, while the largest radius has many hybridized modes, the Rabi splitting is not easily visible in the extinction spectra shown in \autoref{Fig2.H2O}c. This is a result of many different hybridized and non-hybridized modes with different detunings contributing to the total extinction cross-section. For the same reason, although there are two clear peaks in the spectra for a $2.5~\mu$m droplet, it is incorrect to interpret the gap between them as Rabi splitting. Doing so would result in a falsely large Rabi splitting.

The mist is formed by a large number of droplets of various radii. Thus, none of the single-radius spectra matches the experimental $OD$ (light blue dot-dashed line in \autoref{Fig2.H2O}c). To calculate the $OD$ and find the corresponding droplet size distribution, we used Eq. \eqref{eq:OD}. To minimize any potential discrepancies stemming from the Lorentzian approximation (gray dashed lines), we employed the experimental permittivity of water (solid blue line) in our fitting process, particularly noticeable in the largest radii in \autoref{Fig2.H2O}c. The calculated $OD$ that best fits the experiment is shown in \autoref{Fig2.H2O}d, along with the droplet size distribution. A gray dashed line shows the extinction spectrum calculated for the same size distribution but with the oscillator strength of the vibrational modes artificially set to zero (i.e. uncoupled droplets). As for the case of individual droplets, it is important to note that despite the presence of two peaks in \autoref{Fig2.H2O}d, care should be taken not to interpret the energy difference between them as Rabi splitting. Instead, the Rabi splitting is quantified individually for each Mie mode, as depicted in \autoref{Fig2.H2O}a, whereas the two peaks observed in $OD$ result from a complex combination of hybridized and non-hybridized modes across numerous droplets of varying sizes.


\begin{figure}[t!]
\includegraphics[width=0.5\textwidth]{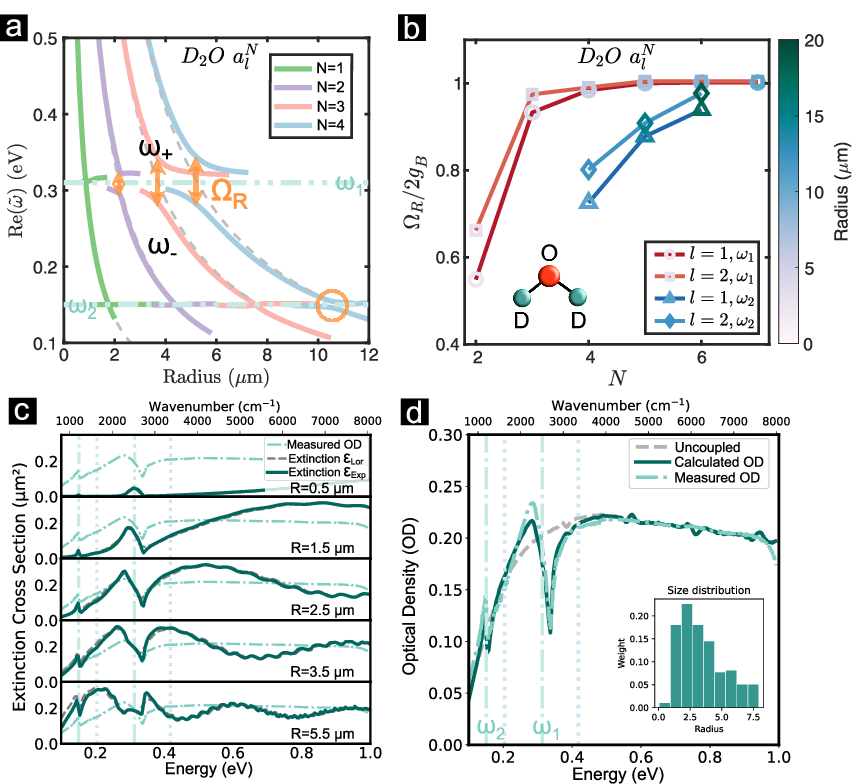}
\caption{\footnotesize{\textbf{Polaritons in D$_{2}$O droplets}. (a) Eigenfrequencies of $a_{l=1}^N$ as a function of droplet radius. In a certain size range, the mode splittings are observed. The Rabi splitting is marked with orange arrows. (b) The Rabi splitting increases with $N$ and converges to the bulk Rabi splitting $2g_B$. (c) Calculated extinction cross-sections of individual water droplets as a function of droplet size. The cross-sections are calculated using experimental (Lorentz) permittivity shown in blue (dashed) lines. The uncoupled molecular vibrations of heavy (regular) water are marked with vertical dash-dotted (dotted) lines. (d) Experimental and fitted optical density using the droplet size distribution shown in the inset. The grey dashed line shows calculated $OD$ for the same size distribution but with vibrational modes artificially switched off, thus representing \emph{uncoupled} spheres. Vertical dash-dotted (dotted) lines mark the position of heavy (regular) water vibrations.
}}
\label{Fig3.D2O}
\end{figure}


To ensure that the observed spectral features for regular water indeed arise from optical and vibrational mode hybridization, we conducted a similar experiment and analysis with heavy water (D$_2$O, 99.9\% Sigma-Aldrich). Although heavy water shares similar optical properties with regular water, its vibrational modes are shifted to lower frequencies due to the increased mass of deuterium in comparison to protium. This red shift corresponds to a factor of $\approx\sqrt{2}$. Hence, we expect that in heavy water droplets, the Rabi splitting appears at different energies and for different droplet sizes. 

In this case, the Lorentzian permittivity was fitted to the experimental values~\cite{Max2009D2OoptCons}. For that the O-D stretch was set to $\omega_{1_{D_2O}}=0.31$ eV, with the same oscillator strength as regular water~\cite{Pastorczak2023oscstrength} $f_{1_{D_2O}}=0.0717$, but with smaller decay rate~\cite{DeMarco2016Losses} $\gamma_{1_{D_2O}}=0.0307$ eV in Eq. \eqref{Eq:Lorentz}. Similarly, the D-O-D bending was set to $\omega_{2_{D_2O}}=0.15$ eV, with the oscillator strength of $f_{2_{D_2O}}=0.0134$ and decay rate of $\gamma_{2_{D_2O}}=0.0076$ eV. 

Like regular water, $a_1^N$ Mie modes exhibit mode splitting for $N\geq2$ as seen in  \autoref{Fig3.D2O}a. However, the Rabi splitting is large enough to enter the strong coupling regime only for $N\geq3$ with radii exceeding $\sim3.7~\mu$m (Figure S5). Larger droplets are therefore necessary to reach strong coupling with heavy water. This is due to the red-shifted vibrational modes of heavy water with respect to regular water.

Despite the larger radii required for strong coupling, the Rabi splitting for heavy water droplets increases for larger $N$, similar to regular water. \autoref{Fig3.D2O}b shows that the Rabi splitting increases until it saturates to the level of bulk Rabi splitting, which in the case of $\omega_{1_{D_2O}}$ is $2g_B \approx 63$ meV. Heavy water also reaches the onset of USC as shown in Figure S6.

After accounting for the contributions of all modes, we obtain the extinction cross-sections for heavy water droplets of various radii (\autoref{Fig3.D2O}c). We observe that the size of the droplet plays a crucial role in determining whether the system enters the strong coupling regime. As for the case of regular water, the two clear peaks observed for $R=2.5~\mu$m cannot be used as a measure of Rabi splitting.

It is important to note that the size distribution of heavy water droplets deviates from that of regular water droplets, owing to variations in physical properties such as ion concentration and viscosity~\cite{Ghazanfari2007mesh}. The inset of \autoref{Fig3.D2O}d displays the size distribution that fits the $OD$ of the heavy water droplets, produced in the same way and conditions as regular water droplets. We observe a larger mean droplet size in the distribution of heavy water compared to regular water. Unsuccessful attempts to fit $OD$ with alternative size distributions are shown in Figure S8. 


This experiment shows that switching regular water to its heavy analog results in a spectrally shifted Rabi splitting because the vibrational modes of heavy water are red-shifted. Moreover, in this case, larger droplets with radii exceeding 3.7 $\mu$m demonstrate strong coupling. This size requirement is notably greater than the minimal size criterion for regular water droplets, which is 2.7 $\mu$m.

The size limit necessary to reach the polaritonic regime in (heavy) water droplets implies that only about half of the droplets are strongly coupled in our laboratory mist realizations. In nature, different types of fogs, mists, and clouds exhibit distinct size distributions of water droplets. For example, a typical marine cloud's mean size is well into the polaritonic regime~\cite{Miles2000heightCloudSize,ZhaoStratusSizePolariton2019}. At the same time, droplets in a fog can be below the polaritonic limit~\cite{Muhammad2007FogCumulusComparison}.


In this work, we employed infrared spectroscopy and Mie theory to measure and analyze the optical density of laboratory mists generated using a mesh nebulizer. We found that the spectra arise from the scattering and absorption of multiple Mie modes, which are strongly coupled to the vibrations of water molecules. We showed that there is a minimal size limit for polaritons to occur. Regular (heavy) water droplets with radii exceeding $\sim 2.7~ \mu$m ($\sim 3.7~ \mu$m) are found to be polaritonic and can even reach the ultrastrong coupling regime. These water droplet sizes are naturally present in mists, fogs, and clouds~\cite{Blau1966CloudSpec,Benayahu1995,Miles2000heightCloudSize}. The polaritonic eigenstates found in these water droplets raise questions about the impact of strong light-matter coupling on their spectroscopic and material properties. Aerosol, atmospheric, and climate researchers may be interested in studying this impact, particularly through monitoring the size of water droplets, the interaction of different aerosols, and the admixture of dissolved molecules to water droplets. Finally, we foresee that droplet polaritons could be realized in non-aqueous liquids.   

\vspace{5mm}

The data supporting the findings of this study are available within the paper and its supplementary material.

We are grateful to Denis G. Baranov and Tomasz J. Antosiewicz for helpful discussions. This work has been supported by the Swedish Research Council (VR Miljö project, grant No: 2016-06059 and VR project, grant No: 2017-04545), the Knut and Alice Wallenberg Foundation (grant No: 2019.0140), Olle Engkvist foundation (grant No: 211-0063) and Chalmers Area of Advance Nano.




\bibliography{Main_water}

\end{document}


\title{Supplemental Material for "Self-hybridized vibrational-Mie polaritons in water droplets"}

\author{Adriana Canales}
\email{adriana.canales@chalmers.se}
\affiliation{Department of Physics, Chalmers University of Technology, 412 96, Göteborg, Sweden.}

\author{Oleg V. Kotov}
\affiliation{Department of Physics, Chalmers University of Technology, 412 96, Göteborg, Sweden.}

\author{Betül Küçüköz}
\affiliation{Department of Physics, Chalmers University of Technology, 412 96, Göteborg, Sweden.}

\author{Timur O. Shegai}
\email{timurs@chalmers.se}
\affiliation{Department of Physics, Chalmers University of Technology, 412 96, Göteborg, Sweden.}

\date{\today}


\maketitle 

\tableofcontents


\newpage

\section{Size distribution of water droplets}

\begin{figure*}[t!]
\centering\includegraphics[width=0.5\textwidth]{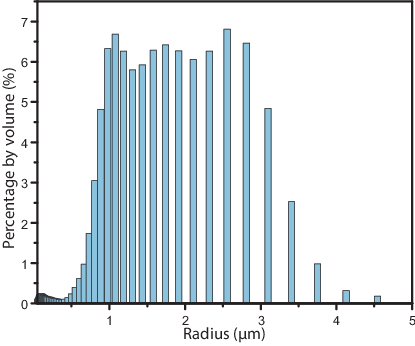}
\caption{{Experimental size distribution of water droplets. Data from Evolu.
}} 
\label{SM.Fig.SizeEvolu}
\end{figure*}

The producing company (Evolu) characterized the size distribution of the water droplets generated by the mesh nebulizer and shared the data with us. The data is presented in \autoref{SM.Fig.SizeEvolu}.

The range of radii was used in the calculations of $OD$, while the percentage by volume guided the initial weights before the fit that resulted in the size distribution presented in Figure 2.

\section{Dispersion of other TE and TM modes}

\begin{figure*}[t!]
\centering\includegraphics[width=0.99\textwidth]{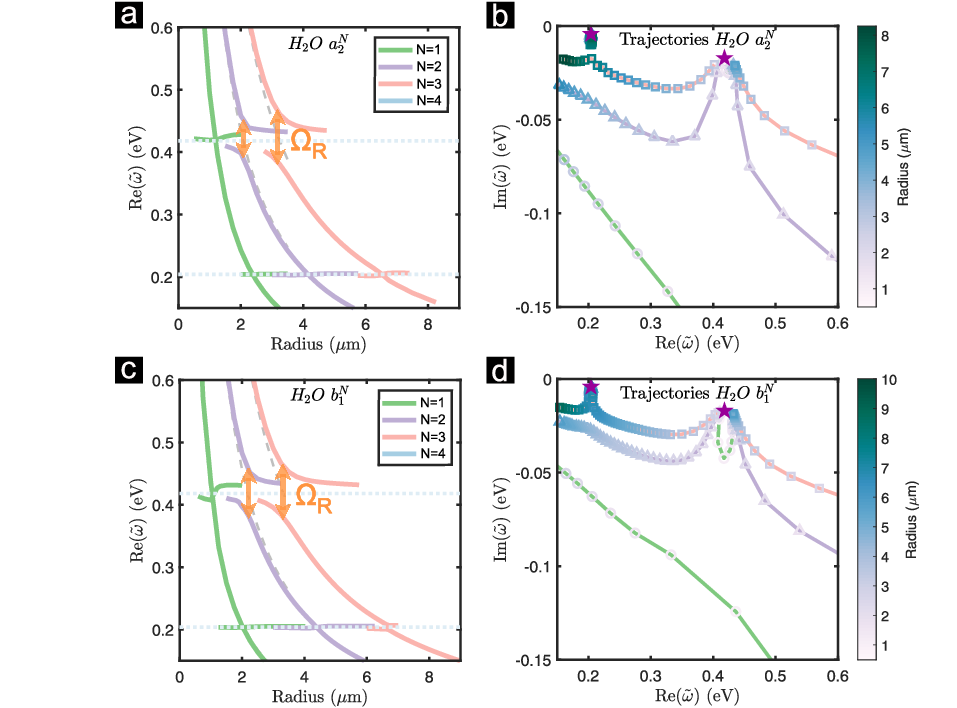}
\caption{{Eigenfrequencies of the system coupled to (top) $a_2^N$ and (c) $b_1^N$. The representation shows only the real part of the eigenfrequencies for increasing radii in (a,c) and the representation in the complex frequency plane in (b,d).  
}} 
\label{SM.Fig.TEandQuad}
\end{figure*}

Different Mie modes are zero-detuned at different radii. As shown in Figure 2b, the Rabi splitting with different modes occurs at different radii. However, at least one Mie mode is always strongly coupled to the vibrational modes for larger radii, even though it may be detuned from the vibrational mode. \autoref{SM.Fig.TEandQuad} shows that $a_2^N$ and $b_1^N$ show strong coupling at zero detuning for different radii.  Therefore, droplets with larger radii have more Mie modes likely to be zero detuned to the vibrational modes. 

The same eigenfrequencies are also plotted in the complex frequency plane in \autoref{SM.Fig.TEandQuad}b,d. The information obtained from the complex frequency plane is outlined below.

\section{Trajectories of the eigenfrequencies in the complex frequency plane}

\begin{figure*}[t!]
\centering\includegraphics[width=0.99\textwidth]{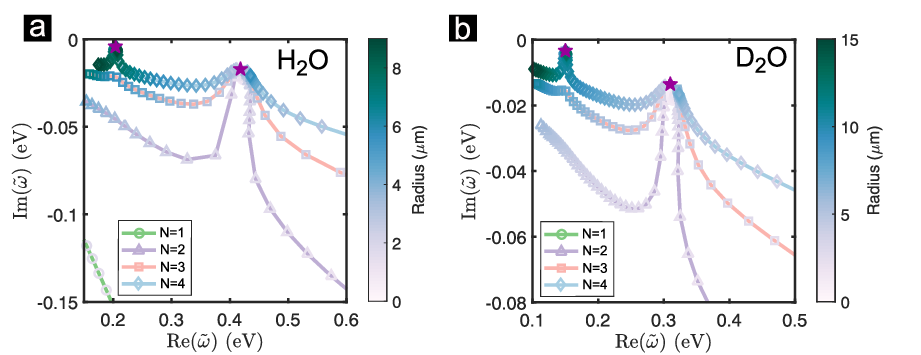}
\caption{{ Eigenfrequencies of (a) water and (b) heavy water droplets coupled to $a_1^N$ in the complex frequency plane. The stars mark the uncoupled vibrational modes.
}} 
\label{SM.Fig.trajectories}
\end{figure*}

In the complex frequency plane, the eigenfrequencies have different topologies in the weak and strong coupling~\cite{SC-canales2021abundance}, as verified here in \autoref{SM.Fig.TEandQuad} and \autoref{SM.Fig.trajectories}. The weakly coupled modes with $N=1$ (green) show a trajectory making a loop close to the vibrational modes (purple stars). The strongly coupled modes split with respect to the real frequency, giving rise to the polaritonic gap around the uncoupled vibrational modes.

As discussed in the main text, the Rabi splitting is compared to the decay rates of the uncoupled components. The decay rate is easily retrieved from the imaginary part of the eigenfrequency. Thus, it is given by the y-axis in the complex frequency plane.

\section{Rabi splitting compared to the uncoupled decay rates}

\begin{figure*}[t!]
\centering\includegraphics[width=0.7
\textwidth]{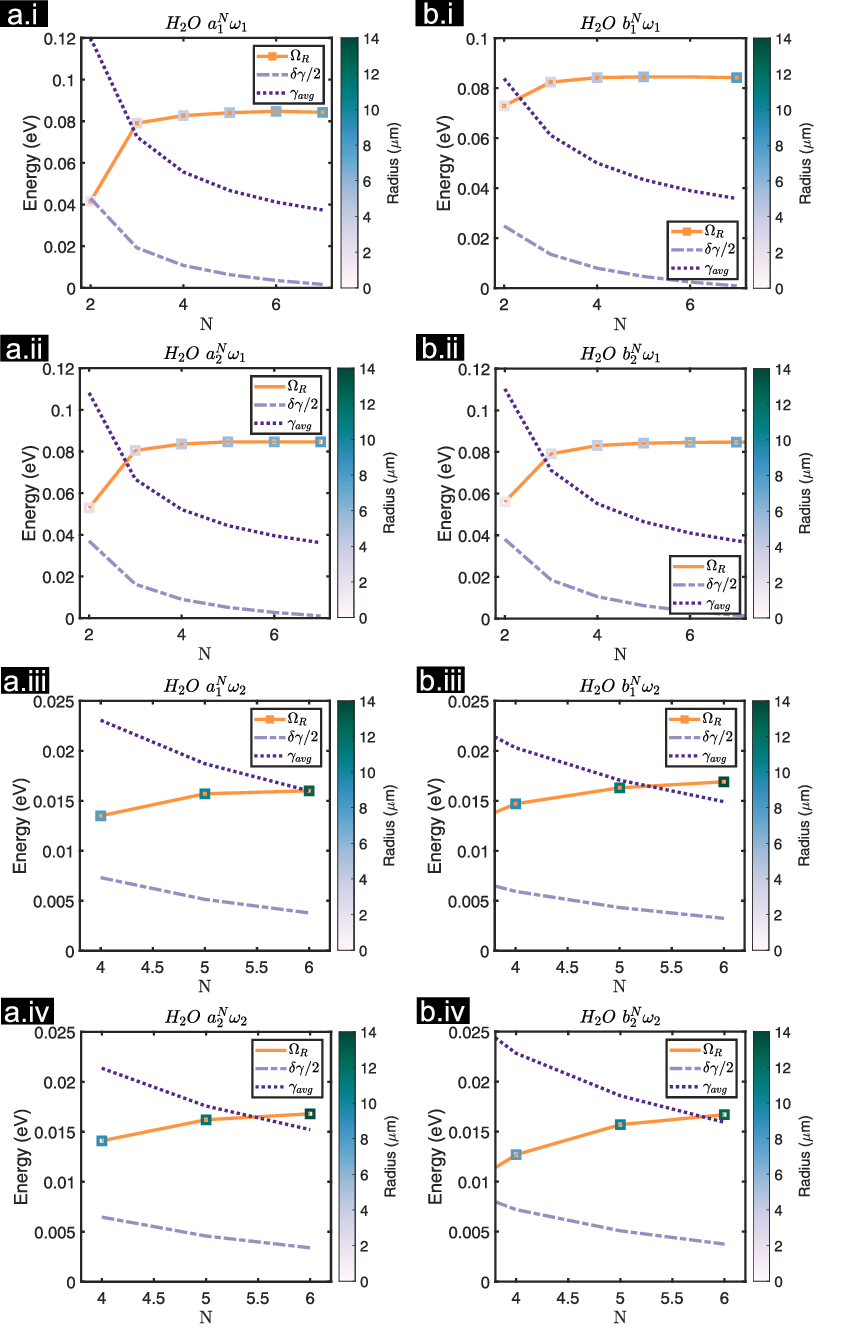}
\caption{{Comparison of Rabi splitting and average loss (dotted line) to verify which Mie modes reach the strong coupling regime with \textbf{regular water} droplets. The exceptional point, after which there is mode splitting, is shown in a dot-dashed line. Column (a) shows the TM modes, while (b) the TE modes. The top two rows describe the coupling with $\omega_1$ with (i) $l=1$ (ii) $l=2$. The last rows show the coupling with $\omega_2$ with (i) $l=1$ (ii) $l=2$.
}} 
\label{SM.Fig.Rabi_loss_H2O}
\end{figure*}

As mentioned above, the definition of strong coupling requires $\Omega_R>\gamma_{avg} = \frac{(\gamma_M + \gamma_v)}{2}$~\cite{torma2014strong}. In \autoref{SM.Fig.Rabi_loss_H2O} we compare the Rabi splitting for several $N$ with $\gamma_{avg}$ and $\delta\gamma/2 = \frac{|\gamma_M-\gamma_v|}{2}$. The latter is the limit after which the exceptional point is passed, and the degenerate modes split. After this point, it is always possible to obtain $\Omega_R$, which is why the purple dot-dashed line is always below the Rabi splitting (orange line) regardless of the Mie mode, $a_l^N,b_l^N$. 

The onset of strong coupling is when the Rabi splitting surpasses the purple dotted line. This happens for $N\geq3$ for all modes in \autoref{SM.Fig.Rabi_loss_H2O} coupling to $\omega_1$, and $N\geq6$ for the ones coupling to $\omega_2$. 

Mie modes with different multipole mode numbers $l$ have an eigenfrequency in resonance to $\omega_1$ for $N=3$ at different radii. The smallest droplet with $\omega_1$ strongly coupled to a Mie mode has a radius of $2.7~\mu$m. In this case, the Mie mode is $b_1^3$ shown in \autoref{SM.Fig.Rabi_loss_H2O}b,i. $a_1^3$ reaches strong coupling for a very similar radius, $2.8~\mu$m.

On the other hand, modes are in resonance with $\omega_2$ only for larger diameters. Therefore, the smallest radius that reaches strong coupling is $12.5~\mu$m, when $b_2^6$ reaches strong coupling with $\omega_2$, as depicted in \autoref{SM.Fig.Rabi_loss_H2O}b,iv. $a_1^6$ reaches strong coupling for $12.6~\mu$m. Thus, droplets with radii larger than that are strongly coupled to $\omega_1$ and $\omega_2$. This is a common radius for water droplets in clouds~\cite{Miles2000heightCloudSize}.

\begin{figure*}[h!]
\centering\includegraphics[width=0.7\textwidth]{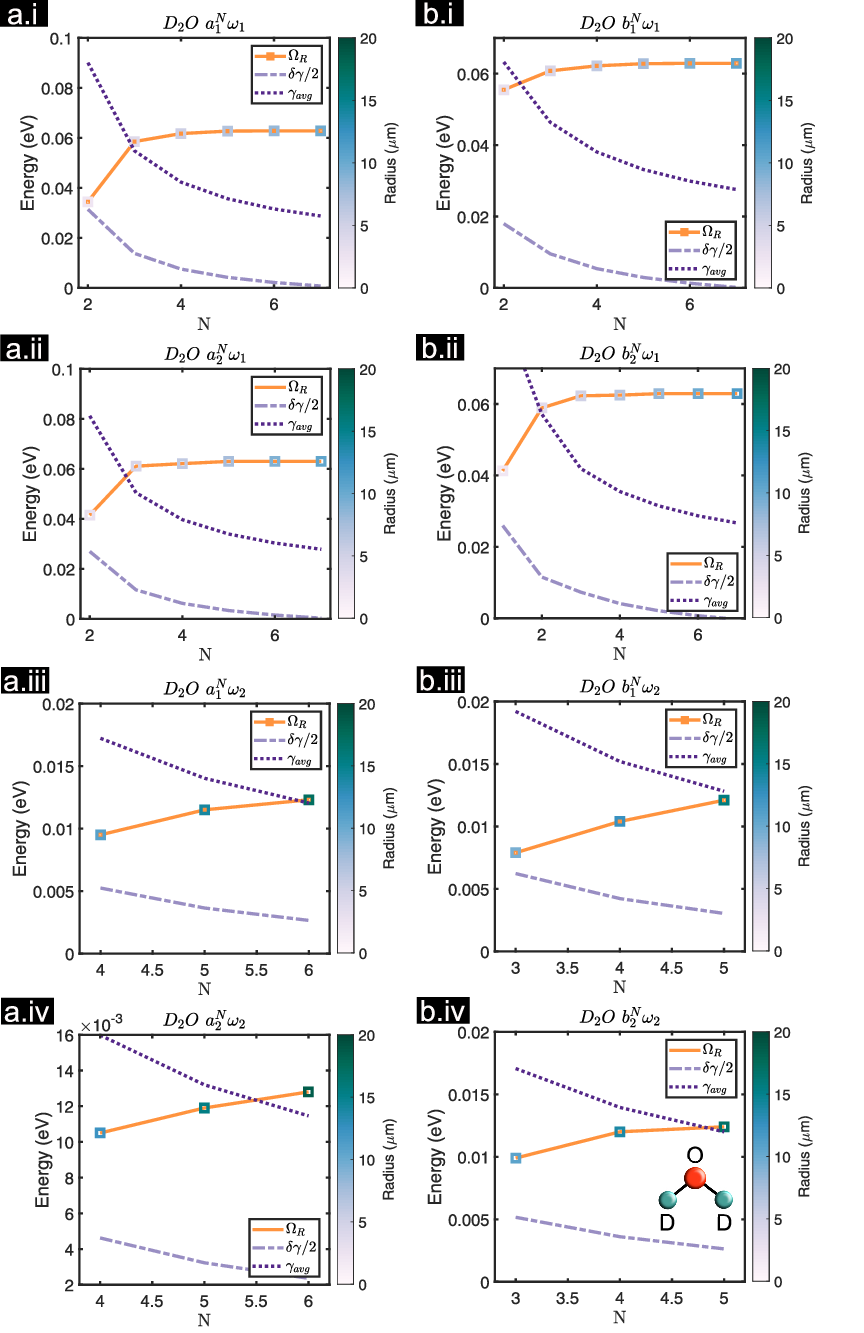}
\caption{{ Comparison of Rabi splitting and average loss (dotted line) to verify which Mie modes reach the strong coupling regime with \textbf{heavy water} droplets. The exceptional point, after which there is mode splitting, is shown in a dot-dashed line. Column (a) shows the TM modes, while (b) the TE modes. The top two rows describe the coupling with $\omega_1$ with (i) $l=1$ (ii) $l=2$. The last rows show the coupling with $\omega_2$ with (i) $l=1$ (ii) $l=2$.
}} 
\label{SM.Fig.Rabi_loss_D2O}
\end{figure*}

The same analysis is done in \autoref{SM.Fig.Rabi_loss_D2O} for heavy water. Similarly, All modes coupled to $\omega_1$ reach the strong coupling regime for $N\geq3$. The modes coupled to $\omega_2$ are almost all in strong coupling after $N\geq6$. However, $b_2^N$ is strongly coupled even after $N=5$.

In the case of heavy water, the smallest radius reaching strong coupling between a Mie mode and $\omega_1$ is $3.7~\mu$m, where the Mie mode is $a_1^1$, as depicted in \autoref{SM.Fig.Rabi_loss_D2O}a,i. The smallest droplet reaching strong coupling to $\omega_2$ is $17~\mu$m. Therefore, larger heavy water droplets are strongly coupled to $\omega_1$ and $\omega_2$.

\section{Normalized Rabi splitting}

\begin{figure*}[h!]
\centering\includegraphics[width=0.9\textwidth]{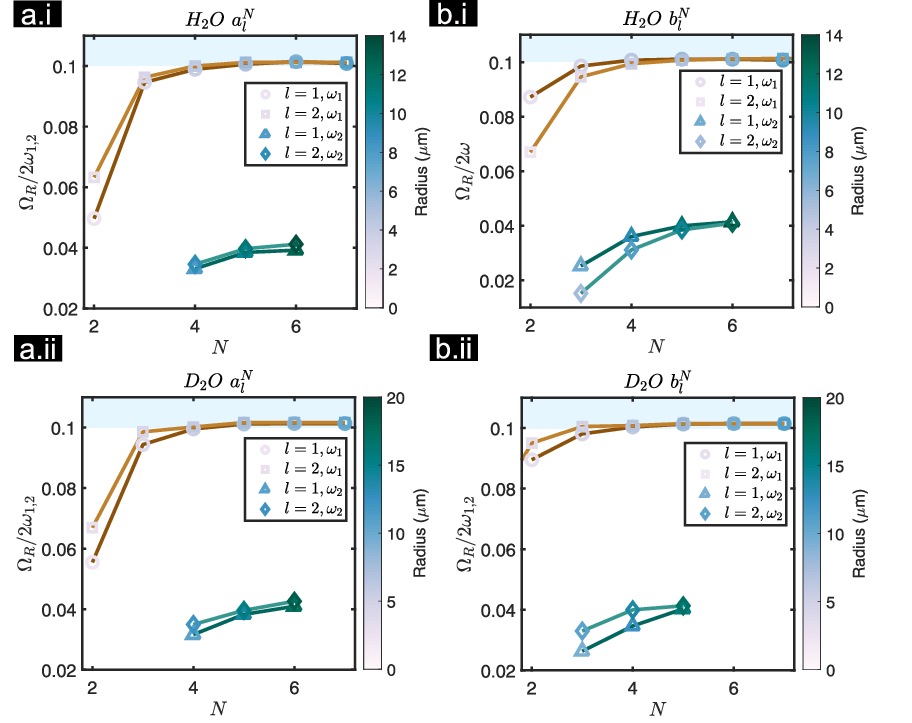}
\caption{{ Normalized coupling strength, $g$ for (a.i) H$_2$O coupled to $a_{l}^{N}$, (b.i) H$_2$O coupled to $b_{l}^{N}$, (a.ii) D$_2$O coupled to $a_{l}^{N}$ and (b.ii) D$_2$O coupled to $b_{l}^{N}$. The ultrastrong coupling regime area is shaded in blue.
}}
\label{SM.Fig.USC}
\end{figure*}

The normalized coupling strength, $g/\omega_{1,2} = \Omega_R/2\omega_{1,2}$, is commonly used to describe the onset of the ultrastrong coupling (USC) regime. The onset is usually set to $g/\omega>0.1$. After this point, higher-order interactions cannot be neglected. 

In \autoref{SM.Fig.USC}, water (top) and heavy water (bottom) reach the onset of USC, particularly with higher radial number modes. Meaning that water droplets can be naturally found in USC regimes.

\section{Optical density of different size distributions of droplets}

\begin{figure*}[t!]
\centering\includegraphics[width=0.95\textwidth]{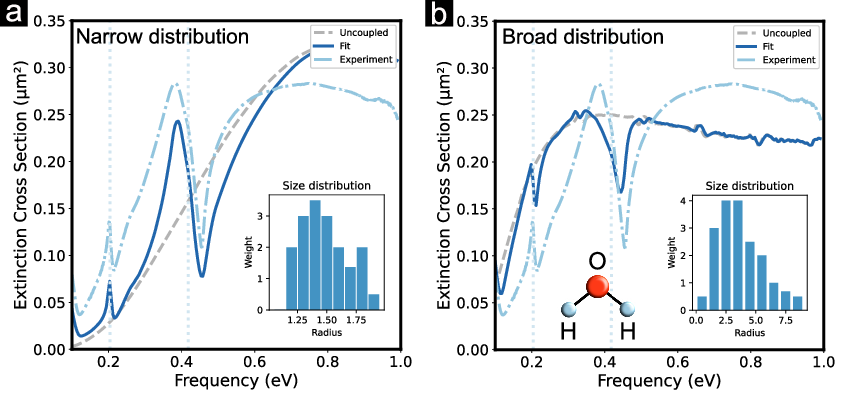}
\caption{{Best fit of total extinction calculations considering different size range of \textbf{water droplets} with a (a) narrower (b) broader distribution. The total extinction of the uncoupled droplets is represented in grey dashed lines, while the uncoupled vibrational modes are plotted in dotted lines. 
}} 
\label{SM.Fig.OtherDist_H2O}
\end{figure*}

In the main text, we show the size distributions best fitting the measured optical density for both water and heavy water. The $OD$ is calculated with various weights for discrete radii to fit the experimental $OD$. The $OD$ is calculated using nine radii in the 0.5-4.5 $\mu$m range. In the case of water, such radii range matches the experimental ones shown in \autoref{SM.Fig.SizeEvolu}. 

However, we tried to fit the experimental $OD$ with various ranges of sizes to check that the range was correct. \autoref{SM.Fig.OtherDist_H2O} shows the best fit obtained for narrow ($R \in [1.1-1.9]~\mu$m) and broad ($R \in [0.5-8.5]~\mu$m) distribution of sizes. Neither case shows similarity to the measured data presented in dot-dashed light blue. The uncoupled spheres $OD$ with the same size distributions are shown in grey dashed lines.

\begin{figure*}[t!]
\centering\includegraphics[width=0.95\textwidth]{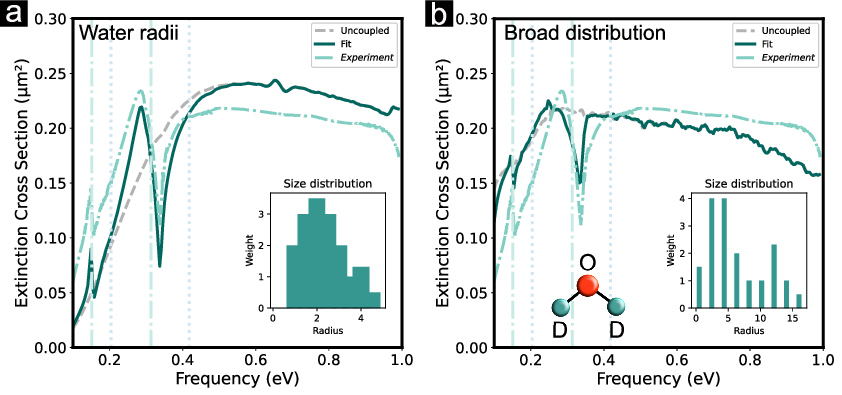}
\caption{{ Best fit of total extinction calculations considering different size range of \textbf{heavy water droplets} with (a) the range of sizes in the case of water droplet. (b) A broader distribution reaching even radii of 15 $\mu$m. The vertical dot-dashed lines represent the uncoupled vibrational modes of heavy water, while the dotted lines represent the water ones.
}} 
\label{SM.Fig.OtherDist_D2O}
\end{figure*}

As discussed in the main text, the physical properties of water and heavy water differ, causing a different size distribution. Therefore, a mismatch is expected between the best $OD$ fit using the radii range for water (0.5-4.5 $\mu$m) and the experimental heavy water $OD$. That mismatch is shown in \autoref{SM.Fig.OtherDist_D2O}a. The actual $OD$ is fit with radii between 0.5 and 8.5 $\mu$m, as used in Figure 3 in the main text. We also tried using a broader range of droplets (0.5-15.5 $\mu$m) also resulted in a large mismatch, as shown in  \autoref{SM.Fig.OtherDist_D2O}b. 

\bibliographystyle{naturemag}
\bibliography{Main_water}